\documentclass[11pt,a4paper]{article}
\usepackage{epsfig}                
\usepackage{amsmath,amssymb,array,dcolumn,subfigure,rotating}
\def\le{\left}
\def\ri{\right}
\def\ln{\operatorname{ln}}
\def\ul#1#2{\textstyle{\frac#1#2}}

\def\dd#1#2{\frac{d^2#1}{{d#2}^2}}

\def\bnabla{\mbox{\boldmath $\nabla $}}
\def\bpsi{\mbox{\boldmath $\psi $}}

\title{Casimir and pseudo-Casimir interactions in confined polyelectrolytes}
\author{R. Podgornik$^{1,2,3}$\footnote{To whom correspondence should be
addressed: rudolf.podgornik@fiz.uni-lj.si} and  J. Dobnikar $^{2,4}$}

\begin{document}
\maketitle

\noindent
$^1$: Laboratory of Physical and Structural Biology, NICHD, Bld. 12A  Rm. 
2041, National Institutes of Health, Bethesda, MD 20892-5626 \\
\noindent
$^2$: Department of Physics, Faculty of Mathematics abd
Physics, University of Ljubljana,
Jadranska 19, 1000 Ljubljana, Slovenia\\
\noindent
$^3$:  Department of Theoretical Physics, J. Stefan
Institute,  Jamova 39, 1000 Ljubljana, Slovenia\\
\noindent
$^4$: Fakult\" at f\" ur Physik, Universit\" at Konstanz,  Pf 5560, D-78457 
Konstanz, Germany\\

\noindent
PACS 30.-v Liquid crystals\\
\noindent
PACS 41.+e Polymers, elastomers and plastics

\begin{abstract}
We investigate the pseudo-Casimir force acting between two charged
surfaces confining a single polyelectrolyte chain with opposite
charge.  We expand the exact free energy to the second order in the
local electrostatic field as well as the replicated polymer density
field around the mean-field (saddle-point) solution.  The quadratic
terms lead to a fluctuation interaction that is partly due to the
(thermal) Casimir effect for the confined electrostatic field and
partly due to the pseudo-Casimir effect due to the confined replicated
polymer density field.  We study the intersurface separation
dependence of both effects and show that the pseudo-Casimir effect
leads to a long range attraction between the surfaces that decays with 
an anomalous algebraic exponent of $\sim 1.7$, smaller than the 
standard exponent of $2$ in the case of Casimir interactions. 
\end{abstract}

\section{Introduction}
Charged polymers have received much attention over the years for their
indisputable importance in industrial applications as well as for
their role in fundamental colloid science (for comprehensive
introductions see \cite{Oosawa,Mandel}).  Apart from the study of
polymers in the bulk, their interactions {\sl with} surfaces, as well
as polymer-mediated interactions {\sl between} surfaces have been
studied on different levels \cite{Napper, Fleer}.  Though
polyelectrolyte interaction with charged surfaces is quite similar to
the general interaction of polymers with neutral surfaces
\cite{Eisen}, the long range nature of the Coulomb potential
introduces additional features and difficulties that make these
problems even more difficult to handle.

\begin{figure}[h]
\begin{center}
        \epsfig{file=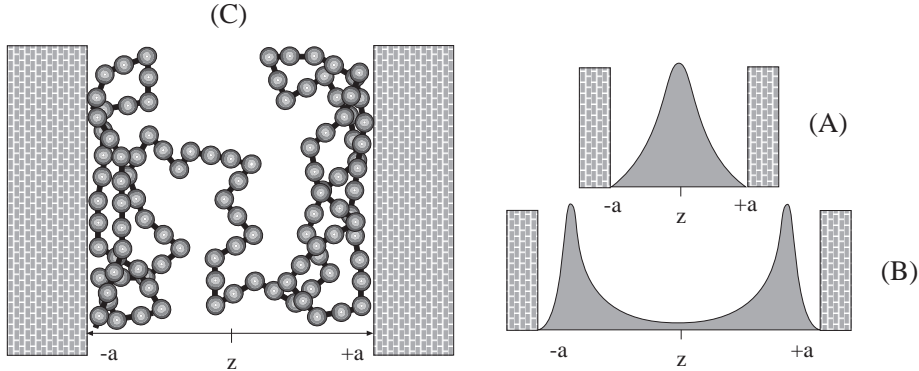, width=5cm, angle=90}
\end{center}
\caption{Schematic representation of the mean-field solution.  For
separations between the surfaces smaller then the critical, the
monomer density distribution is monomodal (A).  For separations larger
then critical the monomer density distribution is bimodal (B),
corresponding to partial adsorption of the chain to the oppositely
charged surfaces.  Parts of the chain that are not adsorbed act as
bridges between the surfaces (C).}
\label{figure0}
\end{figure}

Most of the work on inhomogeneous polyelectrolytes was done on the
mean-field level and the effect of thermal fluctuations has not been
considered at all.  The mean-field picture of the behavior of a
polyelectrolyte chain confined between two oppositely charged surfaces
that emerged from previous studies \cite{Rudi,Andelman} (see Fig. 
\ref{figure0}) lets one think that at small intersurface separations the
polyelectrolyte mediated interations are quite similar to those in the
case of neutral polymers and decay algebraically with separation
\cite{Napper, Fleer}.  At a critical value of the intersurface spacing
there is a transition of the confined polyelectrolyte chain from a
monomodal to a bimodal monomer density distribution.  This means that
at sufficiently large intersurface separations the polymer chain is on
the average partly electrostatically adsorbed to both surfaces and the
remaining part of the chain {\sl bridges} the region between the
surfaces.  The consequence of these {\sl bridging} interactions is an
exponentially decaying attraction between the surfaces that persists
for all separations greater than the critical value
\cite{Rudi,Andelman}.  In this respect the charged chain between
oppositely charged surfaces behaves on the mean-field level a lot like
a neutral chain with a finite adsorption energy to the two bounding
surfaces.

Recent theoretical studies \cite{Netz} however point to conclusion
that thermal fluctuation effects in Coulombic systems, ignored on the
mean-field level, can hardly ever be considered as negligible.  In the
limit of small electrostatic coupling it has been known for a while
that thermal fluctuations in an inhomogeneous Coulomb fluid lead to
zero order van der Waals (or equivalently Casimir) forces
\cite{Rudifaraday,Netz}.  More recent studies of thermal effects in soft
matter in general \cite{Kardar,Krech} are very much consistent with
these findings.  It has been established that thermally driven
fluctuations of order parameters of confined systems also lead to long
range pseudo-Casimir forces \cite{Ajdari}, quite similar in nature to
ordinary Casimir forces except that instead of being due to
electromagnetic field fluctuations they are a consequence of a general
order parameter fluctuations (thus {\sl pseudo}-Casimir interactions). 
In view of this we suspect that similar phenomena should also show up
in the study of fluctuation effects of confined inhomogeneous
polyelectrolytes.

In order to prove this conjecture we write the partition function for
a polyelectrolyte chain confined between two oppositely charged planar
surfaces in such a way that we were able to expand it to quadratic
order in the relevant order parameters, in this case the local
electric field and the replicated polymer density field (see below),
around the mean-field profile and evaluate on this level the effect of
thermal fluctuations.  We note that thermal fluctuations decouple into
two separate contributions: the Casimir contribution due to electric
field fluctuations and the pseudo-Casimir contribution due to the
replicated polymer density field fluctuations.  We show that the
effect of thermal fluctuations depends on only one coupling parameter
$\alpha_{MF}~=~\frac{4\pi}{3}\frac{\sigma \ell}{\tau}$, where $\sigma$
is the surface charge density, $\tau$ is the line charge density of
the polymer chain and $\ell$ is the Kuhn's segment length.  Depending
on the value of this parameter the behavior of the system can be
successfully described on the mean-field level ($\alpha_{MF} \gg 1$)
or has to be ammended by taking into account the fluctuation terms
($\alpha_{MF} \sim 1$).  The most important consequence of the
fluctuation effects seems to be the long range pseudo-Casimir force,
with an anomalous algebraic dependence on the intersurface separations
with an exponent $1.7$, due to the confined fluctuations of the
polymer density field that decays slower than the usual Casimir (or
equivalently van der Waals) force.

The plan of the paper is as follows.  In order to investigate the
fluctuation effects we first write the free energy of a confined
charged polymer chain between two surfaces in the form of an ${\cal
O}(n)$ scalar field theory.  From here the free energy is obtained via
the standard $n \longrightarrow 0$ replica trick \cite{Orland}.  Since
the mean-field is given by the saddle-point, we then expand this field
theory to the second order around the saddle-point in both the
electrostatic field and the replicated polymer density field. 
Calculation of the quadratic corrections to the mean-field free energy
that can be performed exactly gives the Casimir and pseudo-Casimir
interactions mediated by the confined electrostatic and confined
replicated polymer density fields.  We investigate the dependence of
the regularized fluctuation interaction free energy on the
intersurface separation in the limit of small and large spacings
compared to the critical separation.  

\section{Mean-field theory}
As a point of departure we shall take a model confined polyelectrolyte
system, composed of two surfaces with a specified surface charge
density ($\sigma $) and a confined, oppositely charged single
polyelectrolyte chain of length $N \ell$, described with an Edwards
hamiltonian
\begin{equation}
{\cal H}[{\bf r}(s)] = {\textstyle{\frac{3}{2\ell}}} \int_{0}^{N\ell}
\dot{\bf r}^{2} (s) ~ds + \ul12 \int_{0}^{N\ell}\int_{0}^{N\ell}
V({\bf r}(s),{\bf r}(s'))~ds ds'.
\label{eq:1}
\end{equation}
We assume that the intersegment interactions are given by the
unscreened Coulombic form $V({\bf r},{\bf r}') = \tau^{2}/4 \pi
\epsilon\epsilon_{0} ~\vert {\bf r}-{\bf r}' \vert$, where $\tau$ is
the charge per unit length.  This is a bit artificial since we
completely ignore the presence of other mobile ionic species such as
counterions and salt ions, but represents a clean, tractable model system
where fluctuation effects can be studied in detail.

Performing now the Hubbard-Stratonovich transformation for the pair
interaction potential, going to the grand canonical ensemble with a
fixed chemical potential $\mu$ for the monomers and adequatly removing
the closed loop polymer configurations from the partition function
{\sl via} the $n \longrightarrow 0$ limit of an ${\cal O}(n)$ scalar
field theory with the field $\bpsi = (\psi_{1}, \psi_{2} \dots
\psi_{n})$, we remain with the de Gennes - des Cloizeaux
representation for the free energy of a charged polyelectrolyte chain
\cite{Kamien}
\begin{equation}
        {\cal F}(\mu, \beta) = \lim_{n \longrightarrow 0} \ul1n \log{
        {\cal Z}(\mu, \beta)} = {\textstyle \lim_{n \longrightarrow 0}} \ul1n \log{
        \int {\cal D} \phi({\bf r}) {\cal D} \bpsi({\bf r})
        ~\exp{\left( -\beta \int d^{3}{\bf r} ~ {\cal H}[\phi({\bf r}),
        \psi_{i} ({\bf r})]\right)}}
        \label{eq:3}
\end{equation}
where
\begin{equation}
        \beta  {\cal H}[\phi({\bf r}), \psi_{i}({\bf r})] =
        \ul12 \frac{\ell^{2}}{6} ~ \sum_{i} \left(  \bnabla
        \psi_{i} \right)^{2} + \ul12 \mu ~ \sum_{i}\psi_{i}^{2} +
        \ul12 \beta \epsilon\epsilon_{0}\left(\bnabla
        \phi \right)^{2} + i\beta\tau ~ \sum_{i}\phi \psi_{i}^{2}.
        \label{eq:4}
\end{equation}
In writing Eq.  \ref{eq:3} we omitted the determinant ${\rm det}
\left( \beta\epsilon\epsilon_{0} \nabla^{2}\right)$ from the
denominator.  As is usual on the one-loop expansion level, this term
is exactly cancelled by the zero order van der Waals term that one
should add to the final total interaction free energy.  This point is
thoroughly explained in \cite{Roland,Rudifaraday}.  See also below.

The path from here will be to obtain the mean-field solution of the
model defined by Eq.  \ref{eq:3} {\sl via} the saddle-point of the
${\cal O}(n)$ scalar field theory along the lines of \cite{Jug2} and then
to evaluate the contribution of quadratic fluctuations around the
saddle-point to the thermodynamic properties of the system.

Evaluation of the saddle-point from the field hamiltonian Eq.
\ref{eq:4} is trivial.  One notices immediatley that all the
$\psi_{i}$ satisfy the same saddle-point equation and thus $\psi_{i}
\longrightarrow \psi$. With this one can write the dependence of the
hamiltonian Eq. \ref{eq:4} on $n$ explicitly \cite{Emery} and the limit $n
\longrightarrow 0$ can be obtained straightforwardly.  One remains
with
\begin{equation}
        \beta {\cal F}_{0} = -\lim_{n \longrightarrow 0}
        \ul1n \log{{\cal Z}_{0}} = \ul12  \int d^{3}{\bf r}~\left(
        \frac{\ell^{2}}{6} ~\left(  \bnabla \psi \right)^{2} + \mu ~ \psi^{2} +
        \beta \epsilon\epsilon_{0} ~ \left(  \bnabla
        \phi \right)^{2} + i\beta\tau ~ \phi \psi^{2}\right),
        \label{eq:5}
\end{equation}
where we introduced the mean polymer density field $\psi$ so that the
local monomer density is $\rho = \psi^{2}$ and the local electrostatic
mean potential $\varphi = i \phi$.  In the planparallel geometry
considered here all the fields depend only on the transverse
coordinate $z$, and the inhomogeneities are confined to this direction
only $\psi = \psi(z), ~ \varphi = \varphi(z) $.  The saddle-point
equations of the hamiltonian Eq.  \ref{eq:4} can thus be written in
the form
\begin{eqnarray}
        \frac{\ell^{2}}{6} \dd{\psi(z)}z = \mu \psi(z) + 2\beta\tau
        \varphi(z) \psi(z) \quad ,\quad\quad
        - \epsilon\epsilon_{0}\dd{\varphi(z)}z =\tau \psi^{2}(z).
        \label{eq:6}
\end{eqnarray}
These mean-field equations are exactly equivalent to those derived
previously \cite{Rudi,Andelman} if one takes into account that the
polyelectrolyte chain is the only mobile charge in the system (no
counterions and no salt).  The first equation of Eq.  \ref{eq:6} is
the equation for the density field $\psi(z) $ of the polymer in an
inhomogeneous external field $\varphi(z) $ while the second one is a
Poisson - Boltzmann equation for the mean electrostatic potential
$\varphi(z) $ of a charged polymer chain with charge density $\rho(z)
= \tau \psi^{2}(z)$.

The second equation of Eq.  \ref{eq:6} can be solved explicitly and
its solution can be manipulated \cite{Rudi} to yield a limiting form
valid close to the boundaries at $z \simeq \pm a$ where the potential
is largest
\begin{equation}
        \epsilon\epsilon_{0} \varphi (z) = - \tau \int_{-a}^{+a}\vert
        z - z' \vert \psi^{2}(z')~dz' \simeq -\tau \vert  z \vert + {\cal
        O}(z^{2}).
\label{eq:7}
\end{equation}
This expression can be obtained rather straightforwardly by Taylor
expanding the solution around $z = \pm a$.  In the opposite limit, $z
\simeq 0$, we can derive $$\epsilon\epsilon_{0} \varphi (z) \simeq
-\tau \int_{-a}^{+a} \vert u\vert \psi^{2}(u) du - \tau \psi^{2}(0)
z^{2} + {\cal O}(z^{4}).$$  Since the mean-field is largest close to
the boundaries we will use the approximate expression valid strictly
only close to $z = \pm a$ in the whole interval.  This approximation
works extremely well everywhere \cite{Rudi} except close to the origin
where $z$ is small anyway.  The reason for introducing this
additional but inessential approximation is that it also helps in
transforming the first of Eqs.  \ref{eq:6} from a non-linear
Landau-Ginzburg type equation into a linear equation, which can be
solved analytically and explicitly in terms of Airy's functions
\cite{Rudi}.

The electroneutrality demands that at the bounding surfaces ($z = \pm
a$) we have $-\epsilon\epsilon_{0} \frac{\partial \varphi }{\partial
{\bf n}} = \tau \frac{N}{S} \equiv \sigma$, where ${\bf n}$ is the
boundary surface normal.  Because of the impenetrability of the
boundaries to the polymer chain we should also have $\psi (z = \pm a)
= 0$.  The unnormalized solution of the first mean-field equation Eq. 
\ref{eq:6} can now be obtained explicitly in the form
\begin{equation}
        \psi(x) \sim Ai(y_{0} - x) Bi'(y_{0}) - Bi(y_{0} - x) Ai'(y_{0}),
\label{eq:7a}
\end{equation}
where we introduced the dimensionless variables $x =
\lambda_{B}^{1/3}z$, $x_{0} = \lambda_{B}^{1/3}a$,
$y_0=6\mu\lambda_{B}^{-2/3}/\ell^2$, $\lambda_{B} = \frac{12 \beta
\tau^{2} }{\epsilon\epsilon_{0}\ell^{2} }$ and $Ai(x)$ and $Bi(x)$ are
the standard Airy functions.  The dependence $y_0 = y_0(x_{0})$ is
obtained from the vanishing density field boundary conditions at the
two bounding surfaces.  We derive \cite{Rudi} that $y_0(x_0)$ grows
linearly with $x_0$ for large $x_0$, and therefore $u_0=y_0 - x_0$
asymptotically approaches a constant $u_0^{\infty}\approx -2.34$,
while for small $x_0$, $y_0$ is negative and behaves approximately
like $y_0 \approx -(\pi/2x_0)^2$.

The main consequence of the mean-field equations is that for very
small intersurface separations the chain is desorbed with a monomodal
density distribution \cite{Rudi,Andelman}.  In this regime of
intersurface separations the interactions between the surfaces are
repulsive, decaying albebraically with plate separation.  At a
critical value of the intersurface separation $x_{0} = 1.986$ the
electrostatic attraction causes the chain to adsorb to both surfaces
creating a polymer bridge leading in its turn to bridging attraction
between the surfaces.  This attraction, though exponentially small,
persists for all separations between the surfaces larger than the
critical.  The intersurface pressure is thus repulsive for $x_{0} <
1.986$ and attractive for $x_{0} > 1.986$.  For details see
\cite{Rudi}.

\section{Fluctuations}
We now turn to the contribution of fluctuations around the saddle
point to the free energy.  First we expand the hamiltonian Eq.
\ref{eq:4} to the second order in deviations from the mean-fields Eq.
\ref{eq:6}.  Our treatment of the fluctuation effect will thus be
based on a quadratic (one-loop) expansion.  The second functional
derivatives of the field hamiltonian can be assembled into a Hessian
of the fields of Eq.  \ref{eq:4}
\begin{equation}
        {\mbox{\boldmath $\mathbb H $}} = \left[ \begin{array}{cc}
-\frac{\ell^{2} }{6} \nabla^{2} + \mu + 2i\beta\tau
        \phi  &  2i\beta\tau \psi_{n}  \\
2i\beta\tau \psi_{n}  &  - \beta \epsilon\epsilon_{0} \nabla^{2}
\end{array} \right] ,
\label{eq:8}
\end{equation}
where $\varphi$ and $\psi_{n}$ are the solutions of the mean-field
equations Eq.  \ref{eq:6}.  While integrating out the mean replicated
polymer density field and the mean electrostatic field fluctuations we
notice that the functional integral over $\psi_{i}$ fluctuations can
be evaluated directly, giving an explicit dependence of the
fluctuational partition function on $n$.  This is similar to the
general case treated by Emery \cite{Emery}.  The limit $n
\longrightarrow 0$ can now be evaluated explicitly leading to the
following contribution of the fluctuations around the mean-fields in
the quadratic order to the free energy
\begin{eqnarray}
       \beta{\cal F}_{2} &=& \lim_{n \longrightarrow 0} \ul1n \log{{\cal
Z}_{2}} = \ul12 {\rm ln} {{\rm det}\left( - \nabla^{2}\right)} +
        \ul12  {\rm ln} {{\rm det}\left( - \nabla^{2} +
        \frac{6}{\ell^{2}}\mu + 12\frac{\beta\tau}{\ell^{2}}\varphi({\bf
        r})
        \right) } + \nonumber\\
        &+& 12 \frac{(\beta\tau^{2})}{\epsilon\epsilon_{0}\ell^{2}}
\int\int  d^{3}{\bf r} ~d^{3}{\bf r}' ~{\cal
        G}({\bf r},{\bf r}')~{\cal  G}_{0}({\bf r},{\bf r}') ~\psi({\bf
r}) \psi({\bf r}'),
        \label{eq:9}
\end{eqnarray}
where $\varphi({\bf r})$ and $\psi({\bf r})$ are the solutions of the
mean-field equations Eq. \ref{eq:6}. The two Green's
functions ${\cal  G}({\bf r},{\bf r}')$ and ${\cal  G}_{0}({\bf r},{\bf r}')$
are the functional inverses defined as
\begin{eqnarray}
      \left( - \nabla^{2} +
        \frac{6}{\ell^{2}}\mu + 12\frac{\beta\tau}{\ell^{2}}\varphi({\bf
        r}) \right) {\cal  G}({\bf r},{\bf r}') &=& \delta^{3}({\bf
r}-{\bf r}') \nonumber\\
        \left( - \nabla^{2} \right) {\cal G}_{0}({\bf r},{\bf r}') &=&
        \delta^{3}({\bf r}-{\bf r}').
        \label{eq:9a}
\end{eqnarray}
The first term in Eq.  \ref{eq:9} comes from the confined electric
field fluctuations and is in fact the standard zero-order Casimir
interaction energy \cite{Trunov} (see below).  The second and the
third terms are due to the confined polymer density field fluctuations
and thus correspond to the pseudo-Casimir interactions.  The free
energy Eq.  \ref{eq:9} is obtained already {\sl after} the
cancellation referred to in the discussion following Eq.  \ref{eq:4}
and thus represents the total interaction free energy including the
zero order van der Waals contribution.  The latter is evaluated for a slab of
thickness $2a$ with boundaries impenetrable to electrostatic fields.

Let us first deal with the two functional determinants in Eq. 
\ref{eq:9}.  Since the mean-field solution depends only on the
transverse coordinate $z$ we need to investigate the functional
determinant of an operator of the type $-\frac{d^{2}}{dz^{2}} + V(z)$
that can be evaluated in two different ways.  One can first of all
find the eigenvalues $\lambda_{n}$ of this operator with the boundary
condition $f_{n}(z = \pm a) = 0$.  The functional determinant comes
out as the product of these eigenvalues $\Pi_{n}\lambda_{n}$.  The
other approach would take into account the van Vleck identity
\cite{Felsager} involving $f_{0}(z)$, the eigenfunction with zero
eigenvalue and with boundary conditions $f_{0}(z=-a) = 0$ and
$f'_{0}(z=-a) = 1$.  The logarithm of the functional determinant can
thus be written equivalently (up to an irrelevant - in this context -
additive constant) in two different ways \cite{Felsager}
\begin{equation}
        {\rm ln}~{\rm det} \left(  -\frac{d^{2}}{dz^{2}} + V(z) \right) =
        \sum_{n}{\rm ln}\lambda_{n} ={\rm ln}\le [f_{0}(z=a)\ri ].
\label{eq:10}
\end{equation}
In what follows we will use both representations together with the
Fourier decomposition in the $(x,y)$ plane and will find both equally
inadequate when approaching the critical intersurface separation
$x_{0} = 1.986$, either from above or below, where the mean polymer
density field goes through a monomodal to bimodal transition.

Of the first two terms in Eq.  \ref{eq:9} the first one is trivial to
evaluate either {\sl via} the van Vleck relation or the eigenvalue
product representation.  First of all we introduce the Fourier
decomposition in the $(x,y)$ plane with a wave vector ${\bf Q}$, since
we only need to take into account the boundaries in the $z$ direction. 
This leads to the standard zero-order (thermal) Casimir interaction in
the form
\begin{equation}
        \ul12 k_{B}T \sum_{\bf Q}{\rm ln}~{{\rm det}\left(  -
        \frac{d^{2}}{dz^{2}} + Q^{2}\right)} = \frac{S~k_{B}T}{4\pi}
        \int_{0}^{\infty}Q dQ ~{\rm ln}\left( \frac{\cosh{Qa}~
        \sinh{Qa}}{Q}\right),
\label{eq:11}
\end{equation}
where $S$ is the total area of the interacting boundary surfaces.

The second term is trickier.  First of all we again introduce the
Fourier decomposition in the $(x,y)$ plane with a wave vector ${\bf
Q}$ and invoke the approximate solution Eq.  \ref{eq:7}.  Using again
the van Vleck identity we derive
\begin{equation}
        \ul12 k_{B}T  \sum_{\bf Q} {\rm ln}~{{\rm det}\left(
        -\frac{d^{2} }{dz^{2} } + Q^{2}  + \frac{6}{\ell^{2} }\mu +
        12\frac{\beta\tau}{\ell^{2} }\varphi(z) \right) } = \frac{S
~k_{B}T}{4\pi}
        \int_{0}^{\infty} Q dQ ~{\rm ln}~\left( C(Q, a)  S(Q, a)\right),
\label{eq:12}
\end{equation}
where we decomposed the determinant of the operator into two separate
contributions stemming from symmetric and antisymmetric modes
\begin{eqnarray}
        C(Q, z) &=& Bi'(y(Q)) Ai(y(Q) - \lambda_{B}^{1/3}z) - Ai'(y(Q))
Bi(y(Q) - \lambda_{B}^{1/3}z)
        \nonumber\\
        S(Q, z) &=& Bi(y(Q)) Ai(y(Q) - \lambda_{B}^{1/3}z) - Ai(y(Q))
Bi(y(Q) - \lambda_{B}^{1/3}z),
\label{eq:13}
\end{eqnarray}
where $y(Q) = \lambda_{B}^{-2/3} \left( Q^{2} + \frac{6\mu}{\ell^{2} }
\right)$ while $Ai(x)$ and $Bi(x)$ are the standard Airy functions
\cite{Stegun}.  The Airy function solutions are of course a
consequence of the approximation Eq.  \ref{eq:7} for the mean
electrostatic potential.  The free energy Eq.  \ref{eq:12} still needs
to be properly regularized by subtracting the infinite bulk and
surface terms, see below.

In what follows we introduced also these dimensionless variables:
$q^2=\lambda_{B}^{-2/3} Q^2$ and $u(q,x,x_0)=y(q,x_0)-x =q^2+y_0(x_0)
-x=u_0(x_0)+q^2+x_0-x$.  Here $y_0(x_{0})$ is the largest zero of the
symmetric function $C(y_0(x_{0}), x_{0})$ which corresponds exactly to
the zero of the mean-field solution Eq.  \ref{eq:7a} at the two
bounding surfaces \cite{Rudi}.

The third term in Eq.  \ref{eq:9} appears to be the most difficult one
to evaluate but fortunately it is possible to prove that within the
approximation Eq.  \ref{eq:7} its contribution to the fluctuation free
energy ${\cal F}_{2}$ is in fact negligible compared to the other two,
{\sl i.e.} Eqs.  \ref{eq:12} and \ref{eq:13}.  Taking into account the
approximate form for the mean electrostatic potential Eq.  \ref{eq:7}
the last term in Eq.  \ref{eq:9} can be evaluated explicitly leading
to
\begin{equation}
12 \frac{(\beta\tau^{2})}{\epsilon\epsilon_{0}\ell^{2}}
\int\int  d^{3}{\bf r} ~d^{3}{\bf r}' ~{\cal
        G}({\bf r},{\bf r}')~{\cal  G}_{0}({\bf r},{\bf r}') ~\psi({\bf r})
        \psi({\bf r}') =
        \frac{\lambda_{B}}{(\epsilon\epsilon_{0})} \int_{-a}^{+a}\!\!\!
        dz\vert z\vert {\cal G}(z,z',Q=0),
\label{eq:24}
\end{equation}
where we took into account that the mean-field solution $\psi({\bf
r})$ depends only on the transverse coordinate $z$ and that again we
can introduce the Fourier decomposition in the $(x,y)$ plane with a
wave vector ${\bf Q}$.  The magnitude of the coupling contribution Eq. 
\ref{eq:24} turns out to be much smaller than the other two, Eqs. 
\ref{eq:11}, \ref{eq:12}, and can thus be safely ignored in numerical 
computations.

The fluctuation free energy, ${\cal F}_{2}$ which is thus composed
only of the contributions from the fluctuational determinants of the
electrostatic field and the replicated polymer density field, still contains
the divergent bulk and surface parts together with the interaction
terms \cite{Rudifaraday,Trunov}.  ${\cal F}_{2}$ is thus formally
infinite.  This infinity can be regularized by subtracting the
infinite bulk and surface contributions \cite{Trunov} and thus in
forming the regularized fluctuation free energy $${\cal W}_{2}(a) =
{\cal F}_{2}(a) - {\cal F}_{2}^{\infty} (a ).$$  Here ${\cal
F}_{2}^{\infty} (a )$ stands for the form of the fluctuation free
energy for large values of the argument but evaluated at a finite $a$
\cite{Trunov}.

Using now the standard form for the mean-field pressure
\cite{Andelman} and adding to it the fluctuation contribution, one can
write for the total pressure between the boundary surfaces
\begin{eqnarray}
        p(a) &=& p_{MF}+ p_{FL}[\delta\phi] + p_{FL}[\delta\psi] = 
        \nonumber\\
        &=& \ul12 k_{B}T\mu ~ \psi^{2}(z=0) - \frac{k_{B}T}{8 \pi}
        \frac{\zeta (3)}{(2a)^{3} } + p_{FL}(a),
\label{eq:14}
\end{eqnarray}
where the first term on the r.h.s. derives from the mean-field
solution, the second one derives from the Casimir term (stemming from
the electrostatic field fluctuations, first term in Eq.  \ref{eq:9}
and the third one derives from the pseudo-Casimir terms (stemming from
the replicated density field fluctuations, second term in Eq. 
\ref{eq:9} with
\begin{equation}
p_{FL}(a) = - \frac{1}{S} \left(\frac{\partial {\cal W}_{2}(a)}{\partial
(2a)}\right).  
\label{eq:14a}
\end{equation}
The next section deals with the evaluation and regularization of the
fluctuation terms in Eq.  \ref{eq:14}.

\begin{figure}[h]
\begin{center}
        \epsfig{file=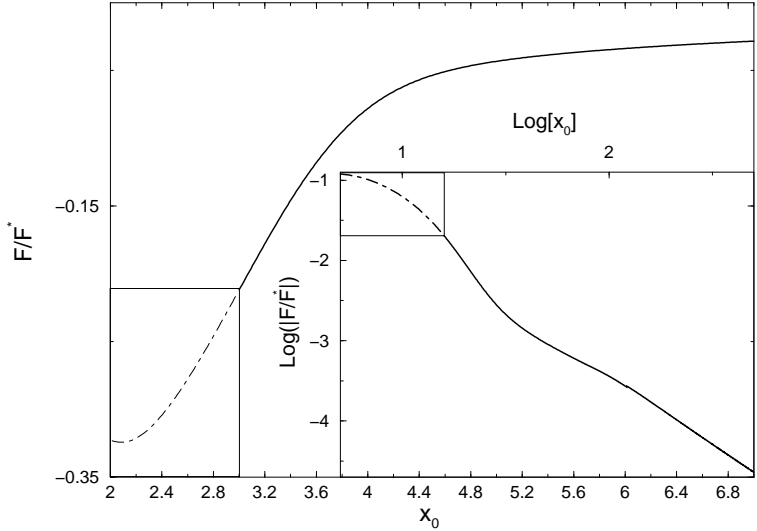, width=7cm, angle=-90}
\end{center}
\caption{The regular part of the fluctuational free energy $F = {\cal
W}_{2}$ in the limit of large separations (see main text) plotted with
respect to the dimensionless separation $x_{0}$.  The region around
the critical separation $x_{0} = 1.986$, where the approximations are
doubtful, is marked with a box.  On the inset there is the same plot
in the logarithmic scale ($\log{\vert \frac{F}{F^*}\vert}$).  One can
read from the inset that the dependence of the free energy on the
separation is algebraic with the exponent around 1,7 (${\cal W}_{2}
\propto x_0^{-1,7}$).  The constant $F^{*}$ is defined as: $F^{*} =
\frac{S kT \lambda_B^{2/3}}{16\pi}$.}
\label{figure1}
\end{figure}

\section{Results - large intersurface separation}
We will first have to regularize the divergent expression Eq. 
\ref{eq:12} by subtracting the divergent terms explicitly in the
limit of large $x_{0}$ (to be quantified below).  When $x_0$ is large
enough the argument $y=u+x_0$ of the Airy functions in Eq. 
\ref{eq:13} becomes large too and we can use the asymptotic expansion
formulas \cite{Stegun} for the Airy functions to explicitly find the
divergent terms in ${\cal F}_{2}^{\infty} (a )$.  The regularized free
energy is then
\begin{eqnarray}
&&{\cal W}_2 (x_0) = \frac{S k_{B}T}{4\pi}\int_{u_0}^{\infty} du
\nonumber\\
&&{\rm ln}\le (
\frac{\le ( Bi'(u+x_0)Ai(u)-Ai'(u+x_0)Bi(u)\ri )\le
(Bi(u+x_0)Ai(u)-Ai(u+x_0)Bi(u)\ri )}
{\le ( Bi'_{L}(u+x_0)Ai(u)-Ai'_{L}(u+x_0)Bi(u)\ri )
\le ( Bi_{L}(u+x_0)Ai(u)-Ai_{L}(u+x_0)Bi(u)\ri )}\ri )\;,
\nonumber\\
~
\label{eq:15}
\end{eqnarray}
where $Ai_{L}$, $Bi_{L}$, $Ai'_{L}$ and $Bi'_{L}$ are the lowest order
asymptotic expressions for the Airy functions \cite{Stegun}.  The next
order in the asymptotic expansion of the Airy function is $Ai_{L2}
(t)\approx -0.07 (\frac{2}{3}t^{3/2})^{-1} Ai_{L} (t)$ and in order
for it to be negligible, $t$ has to be large enough, typically larger
then $1.5$.  This leads to the condition $u_0(x_0)+x_0 > 1.5$, which
yields roughly that $x_0$ has to be larger than $\approx 3$ if our
calculation is to be accurate to within a few percents.  For smaller
values of the intersurface separation the form of the regularized
interaction free energy Eq.  \ref{eq:15} does not work.

Fig.  \ref{figure1} shows the dependence of the dimensionless
fluctuational free energy ${\cal W}_{2}$ on the dimensionless
separation $x_0$.  As already stated the regularization procedure Eq. 
\ref{eq:15} can only be used for sufficiently large intersurface
spacing.  The log-log plot (see the inset of Fig.  \ref{figure1})
reveals that asymptotically the fluctuational contribution with an
algebraic decay prevails.  From the inset of Fig.  \ref{figure1} we
read that  the free energy decays asymptotically as $x_{0}^{- 1.7}$. 
The algebraic decay of the free energy and concommitantly of the
pressure, indicates that the fluctuation induced force has a much
longer range than the mean field force which decays approximately
exponentially \cite{Rudi,Andelman}.  Also it has a longer range than
the zero-order Casimir (or equivalently the zero order van der Waals)
interaction free energy that decays as $x_{0}^{-2}$.

There is also no obvious connection between this result and the
behavior of the fluctuation part of the interaction free energy of a
system composed of unconnected counterions treated on the same level
of approximmation \cite{rudiphys}.  In that case the fluctuation free
energy decays as $x_{0}^{- 2}\ln{x_{0}}$, which is much faster than
$x_{0}^{- 1.7}$.

\section{Results - small intersurface separation}
For small plate separations the above procedure cannot work because in
that case we can not write down explicitly the asymptotic form of the
Airy functions that would be valid in the whole range of integration
in Eq.  \ref{eq:12}.  We found out that the best way to approach the
evaluation of the fluctuation determinant in this limit is via a
perturbation analysis using the eigenvalue representation of the
functional determinant Eq.  \ref{eq:10}.

The perturbative treatment of the evaluation of Eq. \ref{eq:12} can be
introduced as follows. The differential equation we have to solve is
\begin{equation}
\le (-\frac{d^{2}}{dx^{2}} + q^{2}+y_0(x_0) +
\vert x-x_0\vert \ri ) \psi=0 \;.
\label{eq:16}
\end{equation}
The last term can be treated as a linear perturbation potential $V(x)
=\vert x-x_0\vert$.  What we will do now is to evaluate the
determinant of the operator in Eq.  \ref{eq:16} in the form of a
perturbation expansion in $V(x)$ up to and including the first order. 
The order zero of the perturbation expansion of the determinant of the
operator Eq.  \ref{eq:16} gives
\begin{equation}
{\rm det}\le ( -\frac{d^{2}}{dx^{2}} + q^{2}+y_0\ri )=
\psi^{0}_{0}(x_0)=\frac{\sinh
2\sqrt{q^2+y_0}x_0}{2\sqrt{q^2+y_0}}\;.
\label{eq:17}
\end{equation}
The final integration over the wavevector $\bf Q$ has to be divided 
into two parts because $y_0$ is negative, leading to
\begin{equation}
{\cal W}_2^{0}=\frac{S k_{B}T\lambda_{B}^{2/3}}{16\pi x_{0}^2}\Biggl [
\int_{0}^{\pi} wdw\,{\rm ln} \frac{\sin w}{w}
-\frac{\zeta(3)}{4} \Biggr ]\;.
\label{eq:18}
\end{equation}

\begin{figure}[h]
\begin{center}
        \epsfig{file=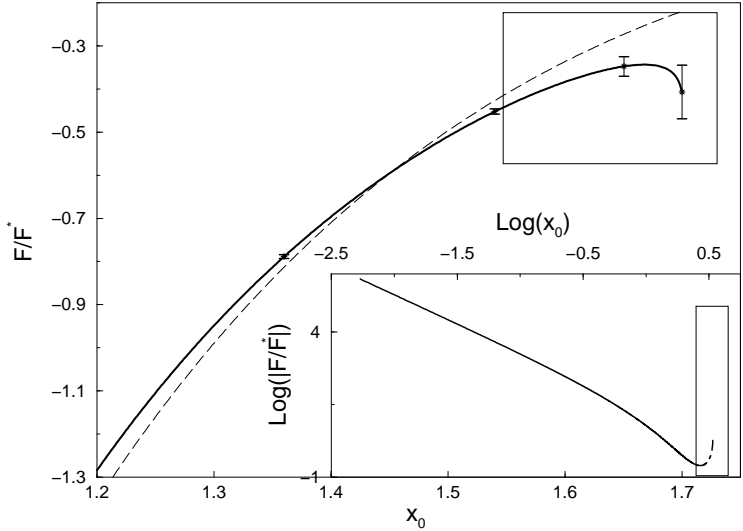, width=7cm, angle=-90}
\end{center}
\caption{The regular part of the fluctuational free energy $F = {\cal
W}_{2}$ in the limit of small separations.  The region around the
critical separation $x_{0} = 1.986$, where the approximations are
doubtful, is again marked with a box.  The zero - order approximation
is shown dashed and the result up to the first order solid.  The error
bars represent the maximal possible error of the first order result. 
The inset represents the log-log plot of the free energy, wherefrom we
deduce that for small separations the fluctuation free energy scales
as $x_{0}^{-2}$.  The constant $F^{*}$ is defined in Fig. 
\ref{figure1}.}
\label{figure2}
\end{figure}

To obtain the first order term in the perturbation expansion we shall
use the product of the eigenvalues representation of the determinant
of the operator which leads to the following first order form of the
free energy
\begin{equation}
{\cal F}_2^{1}=\frac{k_{B}T\lambda_{B}^{2/3}S}{4\pi}
\int_0^{\infty} qdq\,\sum_{N}\,{\rm ln}\lambda^{1}_{N}
\label{eq:19}
\end{equation}
where $\lambda^{1}_{N}$ represents the first order correction to the
eigenvalue of the differential equation Eq. \ref{eq:16}. This 
correction can be evaluated by standard formulas of the perturbation 
theory \cite{Landau} leading to the following result 
\begin{equation}
\lambda^{1}_{N}=\lambda^{0}_{N}-\frac{<N\vert V(x) \vert N>}{<N\vert N>}
=\lambda^{0}_{N}-
\frac{\int_0^{2x_0}\vert x-x_0\vert \sin^2\le (\frac{N\pi}{2x_0}x\ri ) dx}
{\int_0^{2x_0} \sin^2\le (\frac{N\pi}{2x_0}x\ri ) dx}=
\lambda^{0}_{N}+\frac{x_0}{2}\;,
\label{eq:20}
\end{equation}
so the sum of the first order corrections to the log of the
determinant of the operator Eq. \ref{eq:19} is
\begin{equation}
\sum_{N}{\rm ln}\lambda^{1}_{N}\approx \sum_{N}{\rm ln}\lambda^{0}_{N}+
\frac{x_{0}^3}{\pi^2} \sum_{N}\frac{1}{N^2 -1+
(\frac{2x_0}{\pi})^2 (q^2+\frac{x_0}{2})}\;.
\label{eq:21}
\end{equation}
Since the first term in this expression has already been calculated,
we now focus on the second one.  This term is obviously divergent.  If
we calculated the bulk contribution, there would be an integral over
$N$ instead of the discrete sum, which is a consequence of the
boundary conditions.  The regularized free energy is thus
\cite{Trunov} obtained as the difference between the sum and the
integral of the same expression.  After some algebra the result of
this manipulation is
\begin{eqnarray}
{\cal W}^{1}_{2}=\frac{S k_{B}T\lambda_{B}^{2/3} x_0}{16\pi}
\Biggl [ &\ul12& \int_{0}^{\sqrt{1-\frac{2x_{0}^3}{\pi^2}}}
dw \Bigl (\pi\le ( \coth\pi w -\cot\pi w-1\ri )-{\rm
ln}\frac{1+w}{1-w} + 2 {\rm arctan}\frac{1}{w}\Bigr )\,+
\nonumber \\
\, +\, &\ul12&\int_{\sqrt{1-\frac{2x_{0}^3}{\pi^2}}}^{\infty}
dw\Bigl ( \pi\le ( \coth\pi w -1\ri ) +
2 \le ( {\rm arctan}\frac{1}{w} -\frac{1}{w}\ri )\Bigr )\Biggr ].
\label{eq:22}
\end{eqnarray}
Even without really calculating the second order we can nevertheless
say something about its magnitude.  The maximal difference between the
first and the second order perturbation terms can be obtained in 
standard notation as \cite{Landau}
\begin{equation}
\delta \lambda^{1}_{N} \leq \frac{1}{\Delta \lambda}\Biggl [
<N\vert V(x)^2\vert N> - \le (<N\vert V(x)\vert N>\ri )^2\Biggr ],
\label{eq:23}
\end{equation}
where $\Delta \lambda$ is the minimal difference between $\lambda_{N}$
and any other $\lambda_{M}$.  A short calculation then yields
$\frac{\delta\lambda^{1}}{\lambda^{1}} \leq \frac{x_{0}^3}{14.8}$.  As
expected, the reliability of the first order result drops as we move
towards larger $x_0$ {\sl i.e.} towards the mean field phase
transition from bellow.

Fig.  \ref{figure2} shows the first and the second perturbation order
of the regularized fluctuation free energy at small intersurface
separations.  The pressure obviously changes sign for sufficiently
large separation ($x_0$ larger than $\sim 1.68$ on the Fig.  \ref{figure2}).  For small
enough separations the free energy scales approximately as
$x_{0}^{-2}$ which is the same scaling form as for the mean-field
part.  It seems to us that the change in sign of the pressure close to
the transition point $x_0 = 1.986$ is not due entirely to artifacts of
the first order perturbation theory for sufficiently large
intersurface separations, but could concievably survive all higher
order perturbation terms.  Unfortunately all approximation schemes
that we could think of broke down close to the transition point and we
thus have to leave the important question about the behavior of the
fluctuation component of the pressure close to the transition
unanswered.

\begin{figure}[h]
\begin{center}
        \epsfig{file=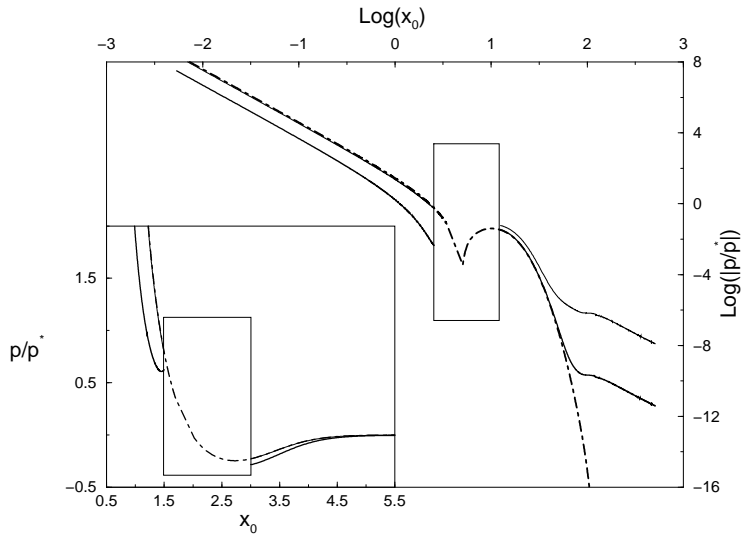, width=7cm, angle=-90}
\end{center}
\caption{The mean field pressure (dot-dashed line) and the total
pressure ($\alpha_{MF} = 4$, thick line and $\alpha_{MF} = 100$, thin
line) - between two plates as a function of the separation in the
logarithmic scale.  The uncertain region around $x_{0} = 1.986$, where
the approximations are doubtful, is marked by a box.  The box contains
only the mean-field results which can be evaluated for the whole range
of $x_{0}$ values.  One can clearly observe the long range tail
stemming from the thermal fluctuations of the monomer density field. 
Inset shows the total pressure - mean field plus fluctuational -
between two plates as a function of the separation for different
values of parameter $\alpha_{MF} = 4,100$.  A large deviation from the
mean field result is obtained for $\alpha_{MF} = 4$ whereas
fluctuations hardly matter for $\alpha_{MF} = 100$.  The constant
$p^{*}$ is defined as: $p^{*} = \frac{\sigma^{2}}{2
\epsilon\epsilon_{0}}$.}
\label{figure3}
\end{figure}

\section{Discussion}
We have described all the contributions to the fluctuational free
energy of the system.  To evaluate the fluctuational part of the
pressure between the two surfaces the derivatives of the regularized
free energy with respect to $x_0$ have been performed numerically. 
The total prerssure can thus be obtained by simply taking the sum of
mean-field and fluctuational contributions Eq.  \ref{eq:14}.  Fig. 
\ref{figure3} shows the dimensionless total pressure as a function of
the dimensionless separation $x_0$ which can be written, Eq. 
\ref{eq:14}, as
\begin{eqnarray}
p(x_0) &=& p_{MF} + p_{FL}[\delta\phi] + p_{FL}[\delta\psi] = \nonumber\\
&=& \frac{k_{B}T }{2}\frac{\lambda_{B} \ell^2 
\sigma}{8~\tau}y_0(x_0)\tilde{\psi}^2(0) - \frac{k_{B}T \zeta(3) 
\lambda_{B}}{64~\pi~x_0^3} - \lambda_{B}^{1/3} 
\frac{d}{d(2 x_0)} \left(\frac{{\cal W}_{2}(x_0)}{S}\right) = 
\nonumber\\
&=& p^* \left[  y_0(x_0)\tilde{\psi}^2(0) - \frac{\zeta(3) 
}{4~\alpha_{MF} ~x_0^3} + \frac{f(x_0)}{ \alpha_{MF} }\right]
\label{eq:25}
\end{eqnarray}
where $f(x_0)$ is a function of $x_0$ only , $\tilde{\psi}(0)$ is the
normalized dimensionless $\psi$, $\alpha_{MF} = \frac{4\pi}{3} \frac{
\sigma \ell }{\tau}$ and $p^* = \frac{k_{B}T \lambda_{B}
\alpha_{MF}}{16 \pi} = \frac{\sigma^2}{2 \epsilon\epsilon_0}$.  The
parameter $\alpha_{MF}$ depends on the charge density on the plates
$\sigma$, the linear charge density on the polyelectrolyte chain
$\tau$ and on the segment length of the polyelectrolyte $\ell$.  If we
put in experimentally reasonable values of $\sigma\approx e_{0}/1,35
nm^2$ then $\alpha_{MF}$ is about 100 for DNA (a relatively stiff
polyelectrolyte) and around 4 for hyaluronic acid (a relatively
flexible polyelectrolyte).  Obviously the larger $\alpha_{MF}$ the
less important will the fluctuation effects be and the more accurate
will be the mean-field result.

Fig.  \ref{figure3} shows the complete pressure Eq.  \ref{eq:25} as a
function of the dimensionless separation for different values of the
coupling parameter $\alpha_{MF}$.  As already stated we were unable to
analyse the behavior of the interaction pressure close to the
mean-field transition point.  Our failure could indicate that close to
the transition point our harmonic approach to the fluctuations breaks
down.  This would not be totally unexpected since we are dealing with
a continuous transition in the distribution of the polyelectrolyte
density that apparently has all the attributes of a second order phase
transition, with large fluctuations close to the transition point.  We
will however not pursue this line of thought here.

It is clear from Fig.  \ref{figure3} that the effect of the
fluctuations on the intersurface force can be profound.  First of all
the fluctuations modify the force at small separations making it less
repulsive if compared to the mean-field case.  Since both the
mean-field as well as the fluctuation contribution to the free energy
scale as $x_{0}^{-2}$ (but with a different sign) it appears that the
effect of the fluctuations is simply to renormalize the magnitude of
the pressure but not its scaling form.  If one would thus try to fit
the complete $p(a)$ with $p_{MF}(a)$ one would have to introduce a
smaller effective charge on the surfaces for the fit to make sense.

At large separations the behavior of $p(a)$ is completely different
from the behavior of $p_{MF}(a)$.  What the fluctuations do is that
they provide a long range algebraic tail to the interactions, which
overwhelms the exponentially screened mean-field attraction at large
intersurface separation.  The algebraic fluctuation tail decays with
an anomalous exponent of $1.7$, that makes the polyelectrolyte
mediated fluctuation interactions even longer ranged than the standard
van der Waals interactions.  This in itself is one of the main
conclusions of our work.  It is also this long ranged tail that is
probably most amenable to experimental observation and verification. 
Since the fluctuation effects are stronger for smaller $\alpha_{MF}$
we predict that they should be easier to measure for very flexible
polyelectrolytes, such as hyaluronic acid as opposed to DNA.

The main drawback of this work is that the fluctuation effects have
been treated on a harmonic level which a priori assumes that they are
small.  However what saves us here is that the mean-field interactions
decay exponentially whereas the fluctuation interactions decay
algebraically.  This is different than in the case of a simple, {\sl
i.e.} unconnected, Coulomb fluid such as an inhomogeneous electrolyte. 
In that case both the mean-field as well as the fluctuation
interactions decay exponentially but the mean-field interactions have
a range twice as long as the fluctuation interactions
\cite{Rudifaraday}.  The case of polyelectrolytes is in this respect
fundamentally different: the range of fluctuation interactions is {\sl
larger} then the range of mean-field interactions and the use of
harmonic approach to fluctuations is probably better grounded in this
case.

\section{Acknowledgements}

The authors would like to thank R. Netz, P.L. Hansen and P. Ziherl
for their comments on an earlier version of the manuscript. We also 
acknowledge the support of the Ministry of Science and Technology 
of Slovenia through the grant \# J1-6168.


\begin{thebibliography}{99}

\bibitem{Oosawa} F.~Oosawa {\sl Polyelectrolytes}, Marcel Dekker, New
York, 1971

\bibitem{Mandel} M.~Mandel {\sl Ency. Polym. Sci. Eng.}, {\bf 11}, 1988

\bibitem{Napper} D.~H.~Napper {\sl Polymeric Stabilization of Colloidal Dispersions},
Marcel Dekker, New York, 1971

\bibitem{Fleer} G.~J.~Fleer, M.~A.~Cohen~Stuart, J.~M.~H.~M.~Scheutjens
{\sl Polymers at Interfaces}, Chapman and Hall, 1993

\bibitem{Eisen} E.~Eisenriegler
{\sl Polymers Near Surfaces : Conformation Properties and Relation to
Critical Phenomena}, World Scientific, Singapore, 1993

\bibitem{Netz} R.~Netz {\sl Lecture notes}, Les Houches \' Ecole de Physique - 
NATO Advanced Study Institute - Euro Summer School {\sl Electrostatic Effects 
in Soft Matter and Biophysics}, Oct. 1-13, (2000) (in preparation)

\bibitem{Rudi} R.~Podgornik {\sl Chem.~Phys.~Lett.}, {\bf 174}, (1990),
191\\
R.~Podgornik {\sl J.~Chem.~Phys.}, {\bf 95}, (1991), 5249\\
R.~Podgornik {\sl Croat.~Chem.~Acta}, {\bf 65}, (1992), 285

\bibitem{Andelman}
I.~Borukhov, D.~Andelman and H.~Orland
{\sl J.~Phys.~Chem.}, {\bf B 103}, (1999), 5042\\
I.~Borukhov, D.~Andelman and H.~Orland 
{\sl Eur.~Phys.~J.~B}, {\bf 5}, (1998), 869\\
I.~Borukhov, D.~Andelman and H.~Orland
{\sl  Macromolecules}, {\bf 5}, (1998), 1665

\bibitem{Joanny}
               J.-L.~Barrat and J.-F.~Joanny
              {\sl Adv.~Chem.~Phys.}, {\bf 94}, (2000), 1-67

\bibitem{Ajdari}
           A.~Ajdari, L.~Peliti and J.~Prost
           {\sl Phys.~Rev.~Lett}, {\bf 66}, (1991), 1481\\
           A.~Ajdari, B.~Duplantier, D.~Hone, L.~Peliti and J.~Prost
           {\sl J.~Phys.~II~France}, {\bf 2}, (1992), 487

\bibitem{Rudifaraday}
           R.~Podgornik and B.~\v Zek\v s
           {\sl J.~Chem.~Soc.~Faraday~Trans.~II}, {\bf 84}, (1988), 611	   

\bibitem{Kardar}
           M.~Kardar and R.~Golestanian
           {\sl Rev~Mod~Phys}, {\bf 71}, (1999), 1233

\bibitem{Krech}
           M.~Krech
           {\sl J~Phys-Condens~Mat}, {\bf 11}, (1999), R391

\bibitem{Orland}
           J.~W.~Negele and H.~Orland
          {\sl Quantum Many-Particle Systems},
               Perseus Press, 1998

\bibitem{Kamien}
               R.~Kamien, L.~Balents, P.~Le~Doussal and E.~Zaslow
              {\sl J.~Phys.~I~France}, {\bf 2}, (1992), 263

\bibitem{Roland}
               R.~Kjellander and S.~Mar\v celja
              {\sl Chem.~Phys.~Letters}, {\bf 142}, (1987), 485
\bibitem{Jug}
               G.~Jug and G.~Rickayzen
              {\sl J.~Phys.~A:~Math.~Gen.}, {\bf 14}, (1981), 1357
\bibitem{Jug2}
               G.~Jug
              {\sl  Ann.~Phys.}, {\bf 142}, (1982), 140

\bibitem{Emery}
               V.~J.~Emery
              {\sl Phys.~Rev.~B}, {\bf 11}, (1975), 239
\bibitem{Trunov}
           V.~M.~Mostepanenko and N.~N.~Trunov
          {\sl The Casimir Effect and its Applications},
           Clarendon Press, Oxford, 1997
\bibitem{Felsager}
               B.~Felsager
              {\sl Particles, Geometry and Fields},
               Springer, Berlin, 1999
\bibitem{Stegun}
               M.~Abramowitz and I.~A.~Stegun
              {\sl Handbook of Mathematical Functions, With Formulas, Graphs, and Mathematical
                     Tables},
               Dover, New York, 1974
\bibitem{rudiphys}
	       R. Podgornik, {\sl J. Phys.  A} {\bf 23} (1989) 275-284 
\bibitem{Landau}
               L.~D.~Landau and E.~M.~Lifshitz
              {\sl Quantum mechanics (Nonrelativistic theory)},
               Pergamom Press, 1977
\bibitem{Casimir}
           H.~B.~G.~Casimir
           {\sl Proc.~Kon.~Ned.~Akad.~Wet.}, {\bf 51}, (1948), 793
\bibitem{KrechK}
           M.~Krech
          {\sl The Casimir effect in critical systems},
               World Scientific, Singapore, 1994
\bibitem{Li}
           H.~Li and M.~Kardar
           {\sl Phys.~Rev.~Lett}, {\bf 67}, (1991), 3275\\
           {\sl Phys.~Rev.~A}, {\bf 46}, (1992), 6490

\end{thebibliography}
\end{document}